\documentclass[10pt,letterpaper]{article}
\usepackage{opex3}
\usepackage{amsmath}
\usepackage{subfigure}

\begin{document}

\title{Dynamic optical lattices: two-dimensional rotating and accordion lattices for ultracold atoms}

\author{R. A. Williams, J. D. Pillet, S. Al-Assam, B. Fletcher, M. Shotter and\linebreak C. J. Foot}

\address{Clarendon Laboratory, University of Oxford, Parks Road, Oxford OX1 3PU, United Kingdom}

\email{r.williams1@physics.ox.ac.uk} 



\begin{abstract}We demonstrate a novel experimental arrangement which can rotate a 2D optical lattice at frequencies up to several kilohertz.  
Ultracold atoms in such a rotating lattice can be used for the direct quantum simulation of strongly correlated systems under large effective magnetic 
fields, allowing investigation of phenomena such as the fractional quantum Hall effect.  Our arrangement also allows the periodicity of a 2D optical 
lattice to be varied dynamically, producing a 2D accordion lattice.
\end{abstract}



\section{Introduction}

Ultracold atoms in optical lattices present a unique system with which to investigate strongly correlated many-body physics with unprecedented 
experimental control.  Direct quantum simulation of many-body Hamiltonians is of fundamental interest in studying the dynamic behaviour of various 
condensed matter systems, such as Hubbard and spin models, disordered systems and high temperature superconductivity \cite{lewenstein07}.  Some of the 
most exotic strongly correlated phenomena in condensed matter physics occur when systems are subjected to large magnetic fields, for example the 
fractional quantum Hall effect \cite{jain}.  There is a close analogy between the physics of electron gases under large magnetic fields and rapidly 
rotating neutral atoms \cite{viefers08}.  Recently it has been predicted that strongly correlated 2D systems under large magnetic fields can be 
simulated using ultracold atoms in a rotating optical lattice, allowing an investigation of high field fractional quantum Hall physics 
\cite{sorenson05, palmer06, bhat07}.  Other interesting phenomena which can be explored with cold atoms in a rotating lattice include the Hofstadter 
butterfly \cite{jaksch2003}, fully frustrated Josephson-junction arrays \cite{polini05, kasamatsu08}, and the dynamics of a vortex array in the 
presence of a co-rotating lattice \cite{tung07, kasamatsu06}.

Only one group has previously made a rotating lattice, where a mechanically rotated mask was rotated at $\sim 8\mathrm{Hz}$ \cite{tung07}.  In this 
paper we present a novel experimental arrangement which allows a 2D lattice to be rotated smoothly at any frequency from 0 to $\sim 1$kHz.  Ultracold 
atoms in a lattice of period $\sim 1\mu\mathrm{m}$ rotating at $\sim 1$kHz experience an effective magnetic flux density of order of one magnetic flux 
quantum per lattice cell.  These high rotation frequencies will allow high field regimes to be explored, where the exotic quantum phenomena discussed 
above are predicted to occur.  The rotation is generated by a dual-axis acousto-optic deflector (AOD) and thus does not suffer from mechanical 
vibrations and instabilities which would lead to significant heating of atoms in the lattice.  The limit of 1kHz in the current apparatus was set by 
the update rate of the radio frequency synthesis system controlling the AOD, however the optical arrangement can accommodate even higher frequencies.

In addition the arrangement allows for the realization of a 2D accordion lattice, that is, an optical lattice with a dynamically variable periodicity.  
1D accordion lattices have previously been demonstrated by a number of groups \cite{huckans06, fallani05, li08}.  Experiments probing the 
superfluid-Mott insulator transition typically use optical lattices with periodicities $\sim 400-500\mathrm{nm}$ where tunnelling between lattice 
sites is significant \cite{greiner02}.  Direct observation of the in-lattice density distribution of a number-squeezed state has not been achieved due 
to difficulties in achieving an optical system with sufficient resolution.  However an accordion lattice could allow for the expansion of the lattice 
spacing to above the resolution of an optical imaging system ($> 1\mu\mathrm{m}$) after the Mott insulator had been achieved.  Detecting and 
manipulating atoms at single lattice sites are important criteria for quantum information processing schemes with neutral atoms \cite{jaksch04}.

\section{Experimental arrangement}

\begin{figure}[t]
\centering
\includegraphics[scale=0.38]{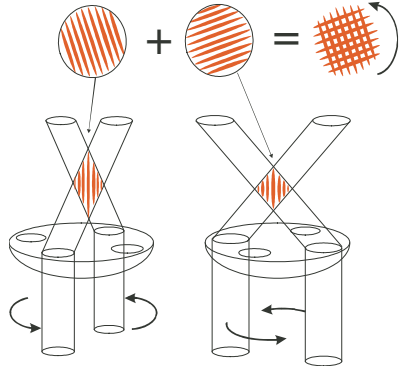}
\caption{Two orthogonal standing wave intensity patterns in the focal plane of a lens combine to form a 2D lattice.  The axial symmetry of the system 
allows rotation of the lattice to be realised.}
\label{fig:rotatinglattice}
\end{figure}

Optical lattices are formed by standing wave intensity patterns of laser light.  While it is common in ultracold atom experiments to use interfering 
counterpropagating laser beams to produce optical lattices with spacings of $\lambda\slash 2$ another often-used technique is having two beams 
intersecting at an angle of $2\theta$, producing a lattice with a spacing of $d = \lambda\slash(2\sin\theta)$.  Two parallel beams incident on a lens 
will form a 1D optical lattice in the focal plane of the lens where they intersect.  Such an arrangement was used recently to realize a 1D accordion 
lattice \cite{li08}. The cylindrical symmetry of such an arrangement allowed us to achieve a rotating lattice upon rotation of the parallel beams 
incident on a lens.  Two orthogonal 1D lattices were combined to form a 2D rotating lattice as shown in Fig. \ref{fig:rotatinglattice}.

\begin{figure}[htbp]
\centering
\includegraphics[scale=0.38]{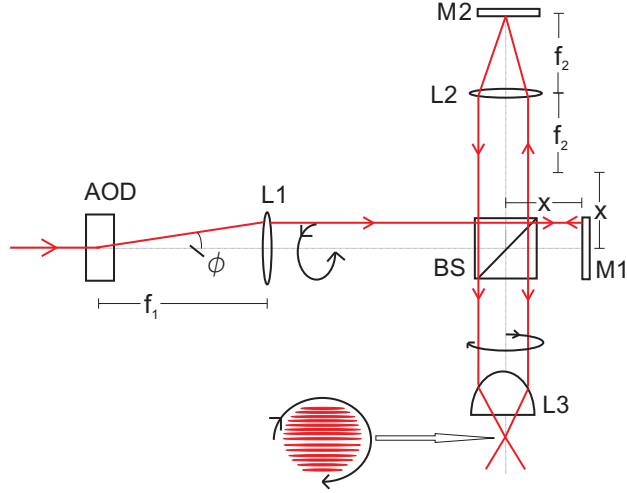}
\caption{Experimental arrangement for the generation of rotating interference fringes.  BS is a non-polarising beam splitter cube, L1-L3 are 
converging lenses, M1 and M2 are mirrors.}
\label{fig:interferometer}
\end{figure}

Figure \ref{fig:interferometer} shows the arrangement used to generate two parallel beams of light which rotated $180^\circ$ out of phase with each 
other about the optical axis of the system.  Clearly an important criterion for a pair of beams is that they must have the same frequency and 
polarization in order to interfere with one another.  The Michelson interferometer-like arrangement shown in Fig. \ref{fig:interferometer} generates 
two output beams from a single input beam.  The long arm of the device contained a lens, L2, which introduced a parallel displacement about the 
optical axis between the two beams.  By generating this second, displaced beam from the first we ensure both beams have the same polarization and 
frequency and thus will interfere with each other.  This would not be the case if two individual AODs were used to rotate two independent beams.  The 
entire setup had cylindrical symmetry about the optical axis of the system and hence rotation could be realized.

The distances between the non-polarizing beam splitter cube (BS), L2 and M2 in the long arm were chosen such that the output beams from the short and 
long arms had the same waist and convergence despite the unequal path lengths.  This can be seen most easily by realizing that the long arm of the 
arrangement in Fig. \ref{fig:interferometer} contains a conventional 4f optical imaging system, that is, a one-to-one telescope, mapping the waist and 
wavefront curvature of the incoming beam onto the beam leaving the long arm \cite{saleh}.  The beam incident on the BS did not therefore need to be 
collimated to ensure that the two output beams had identical Gaussian characteristics.  The unequal path lengths of the different arms mean that the 
laser must have a coherence length greater than the path difference, however this requirement is easily satisfied by lasers typically used for 
creating optical lattices and by a reasonable choice for L2.  For L2 we used a 10cm focal length achromatic doublet, while a Titanium-doped Sapphire 
laser operating at 830nm (coherence length $\gg 1\mathrm{m}$) was used for the lattice beams.

\begin{figure}[t]
\centering
\includegraphics[scale=0.4]{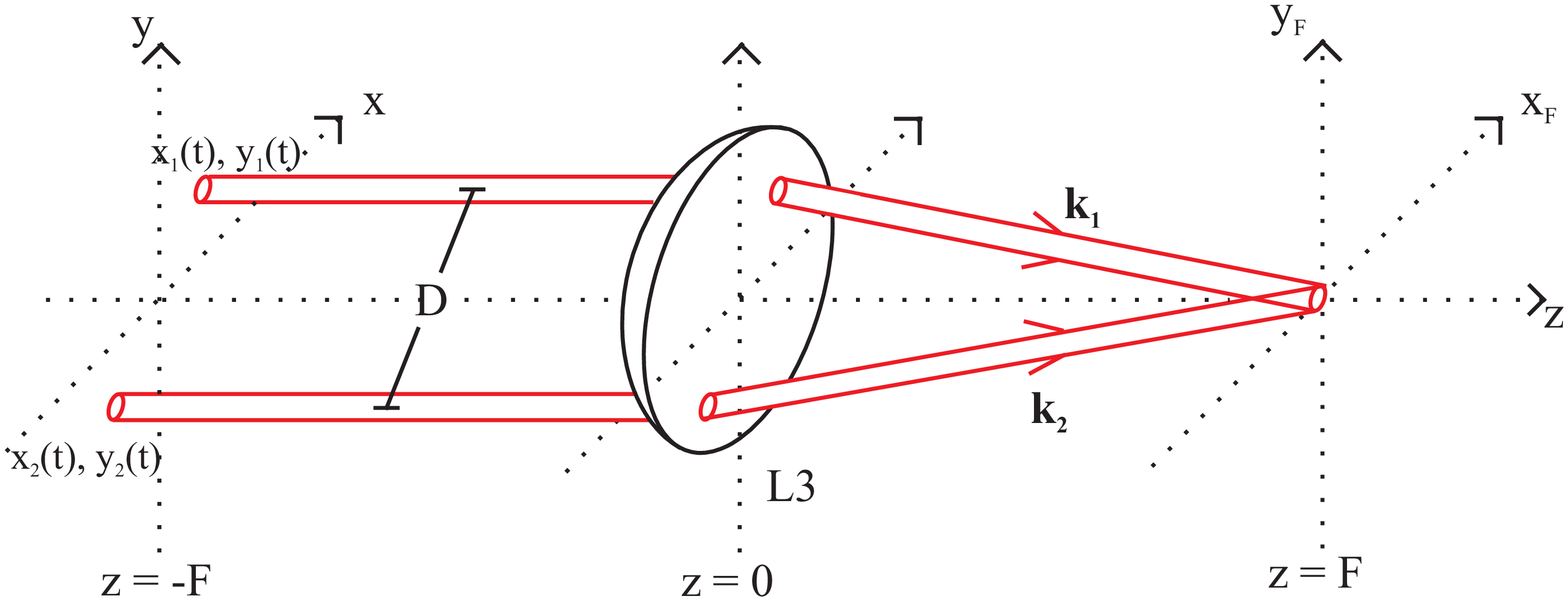}
\caption{It is convenient to describe the resultant intensity in the back focal plane of L3 ($z=F$) in terms of the coordinates of the incident beams 
in the front focal plane of the lens ($z=-F$).  The arrangement of Fig. \ref{fig:interferometer} produced the second beam such that $x_2(t) = 
-x_1(t)$, $y_2(t) = -y_1(t)$, while $x_1(t)$, $y_1(t)$ were controlled by steering the laser beam with the AOD.  $D$ is the separation of the two 
beams incident on L3.}
\label{fig:frontbackfocalplane}
\end{figure}

We relate the electric field distributions in the front and back focal of L3 using the geometrical optics argument shown in Fig. 
\ref{fig:frontbackfocalplane}, considering both the ideal case of a thin lens and the more realistic case of a high numerical aperture objective lens.  
Neglecting the Gaussian envelope of the beams the electric field in the back focal plane of L3 is
\begin{eqnarray}
\mathbf{E}(\mathbf{r}_F,F) &=& \mathbf{E}_0[\cos(\mathbf{k}_1\cdot \mathbf{r}_F - \omega t) + \cos(\mathbf{k}_2\cdot \mathbf{r}_F - \omega t)]\\
&=& 2\mathbf{E}_0\cos\left(\frac{(\mathbf{k}_1+\mathbf{k}_2)\cdot \mathbf{r}_F}{2} - \omega t\right)\cos\left(\frac{(\mathbf{k}_1-\mathbf{k}_2)\cdot 
\mathbf{r}_F}{2}\right)
\label{eqn:electricfields}
\end{eqnarray}
where $\mathbf{r}_F$ designates coordinates in the back focal plane of L3, $F$ is the focal length of lens L3, $\mathbf{k}_1$ and $\mathbf{k}_2$ are 
the wavevectors of the intersecting beams as shown in Fig. \ref{fig:frontbackfocalplane} and $\omega$ is the angular frequency of the laser beams.  
The resultant intensity thus has the form  
\begin{equation}
I(\mathbf{r}_F,F) = 2I_0[1 + \cos(\mathbf{k}_1-\mathbf{k}_2)\cdot \mathbf{r}_F].
\label{eqn:intensityone}
\end{equation}

Equation \ref{eqn:electricfields} considers two beams with identical linear polarization parallel to the resultant intensity fringes.  While it is 
always possible to arrange the beam polarizations like this for a two-dimensional accordion lattice this is not the case for a rotating lattice.  If 
using linear polarization with a rotating beam the electric field component in the plane of the lattice will change with rotation, leading to a 
modulation in lattice depth.  This problem can be eliminated by using the same circular polarization for a pair of interfering beams.  Given the 
symmetry of the arrangement the electric field component in the plane of the interference fringes remains constant as the lattice rotates, and there 
is no modulation of the lattice depth.

From the geometry shown in Fig. \ref{fig:frontbackfocalplane} (assuming a thin lens) wavevectors $\mathbf{k}_1$, $\mathbf{k}_2$ can be expressed as 
$\mathbf{k}_i = (2\pi\slash\lambda)(-x_i\mathbf{e}_x-y_i\mathbf{e}_y+F\mathbf{e}_z)\slash(x_i^2+y_i^2+F^2)^{1/2}$, where $x_i(t),y_i(t)$ are the 
coordinates of the beams in the front focal plane of L3.  However high numerical aperture objective lenses have a finite thickness and are designed to 
obey the Sine Condition \cite{born} to ensure good imaging of off-axis objects.  A consequence of obeying the Sine Condition is that the wavevectors 
$\mathbf{k}_1$, $\mathbf{k}_2$ make a different angle to the optical axis, with $\mathbf{k}_i = 
(2\pi\slash\lambda)(-x_i\mathbf{e}_x-y_i\mathbf{e}_y+F\mathbf{e}_z)\slash F$.  The optical system of Fig. \ref{fig:interferometer} created two beams 
in the front focal plane of L3 such that $x_2(t) = -x_1(t)$, $y_2(t) = -y_1(t)$.  Equation \ref{eqn:intensityone} then becomes
\begin{equation}
I(x_F,y_F,F) = 2I_0[1 + \cos\frac{2\pi}{\lambda}\left(\frac{2x_1x_F+2y_1y_F}{F}\right)].
\label{eqn:intensitytwo}
\end{equation}
We controlled $x_1(t)$ and $y_1(t)$ by steering the beam with the AOD, e.g.\ for a lattice rotating at an angular frequency of $\Omega$, $x_1(t) = 
(D\slash 2)\cos\Omega t$, $y_1(t) = (D\slash 2)\sin\Omega t$, where $D$ is the separation of the two beams in the front focal plane of L3.  
Substituting this into Equation \ref{eqn:intensitytwo} one finds 
\begin{equation}
I(x_F,y_F,F) = 2I_0[1 + \cos\frac{2\pi}{d}(x_F\cos\Omega t+y_F\sin\Omega t)]
\label{eqn:intensitythree}
\end{equation}
where $d = \lambda F\slash D$ is the periodicity of the lattice.  Li {\it{et al}} \cite{li08}  measured the periodicity of lattices formed in the 
focal plane of an achromatic doublet and found this expression for the lattice spacing to be accurate for intersection angles up to $38^\circ$.  
Equation \ref{eqn:intensitythree} describes a one-dimensional lattice rotating with angular frequency $\Omega$.

An accordion lattice can also be implemented in this arrangement by using the AOD to change $D$, the separation of the beams in the front focal plane 
of L3.  $D$ is determined by the angle of deflection of the beam from the optical axis ($\phi$ in Fig. \ref{fig:interferometer}) and the focal length 
of L1, $D = 2f_1\tan\phi$.  The smallest lattice spacing achievable, $d_\mathrm{min}$, is set by the maximum intersection angle of the beams and hence 
the numerical aperture (N.A.) of objective lens L3, $d_\mathrm{min} = \lambda\slash (2\mathrm{N.A.})$.  The minimum lattice periodicity possible is of 
considerable importance in experiments with cold atoms as quantum tunnelling and the mutual interaction energy of two atoms on the same site become 
negligible at larger lattice spacings.  High N.A. objective lenses have been successfully interfaced with cold atoms, for example in 
\cite{schlosser01} an N.A. of 0.7 was used.  For 830nm light this would give $d_\mathrm{min}\approx 600$nm.  The largest possible lattice period was 
limited by the Gaussian profile of the laser beams, that is, to achieve a reasonable number of fringes in the central, uniform intensity region of the 
beam profile $d$ is required to be smaller than the beam waist.

The generation of rotation of the initial single beam was produced using a dual axis AOD (an Isomet LS110A-830XY).  The center frequency of the AOD 
was 50MHz, with a bandwidth of 25MHz.  The radio frequency (rf) signal to the AOD controlling the deflection was generated using a custom-made Direct 
Digital Synthesis (DDS) based system, providing powerful and fast control of the beam deflection.  AODs are specifically optimised for deflecting 
laser beams, giving large deflection angles and a reasonably uniform diffraction efficiency over their bandwidth.  However some power variation in the 
deflected beam remained, which could be as much as $\pm 5\%$ in our arrangement.  Such power fluctuations are undesirable when implementing an optical 
lattice, and we were able to compensate for the deterministic change in diffraction efficiency with deflection angle by calibrating the rf wave 
amplitude appropriately at each frequency.  Alternatively a feed forward mechanism could be used to correct for the power variation as was done in 
\cite{schnelle08}, or for low rotation frequencies the beam power after the AOD could be actively stabilized using a feedback arrangement.

While Fig. \ref{fig:interferometer} shows the arrangement for generating a single set of rotating fringes it is straightforward to combine two 
orthogonal sets of fringes to create a two-dimensional optical lattice.  The most straightforward and effective way to do this is to introduce a 
second AOD to rotate a beam $\pi\slash 2$ out of phase with the first.  These two rotating beams can then be combined at a beam splitter cube before 
both entering the Michelson interferometer-like optics.  In order to prevent interference between the two one-dimensional lattices one pair of lattice 
beams must be detuned by MHz relative to the other pair.  This can be achieved by operating the two AODs about different center frequencies which are 
separated by MHz.  In this case the rf signal applied to each AOD is modulated sinusoidally, with the center frequency for each AOD separated by a 
frequency larger than twice the modulation amplitude.  Alternatively an acoustic-optic modulator can be used to shift the frequency of a beam before 
it enters an AOD.

\section{Demonstration of two-dimensional rotating and accordion lattices}

\begin{figure}[htbp]
\centering
\subfigure[]{\label{fig:30}\includegraphics[scale=0.35]{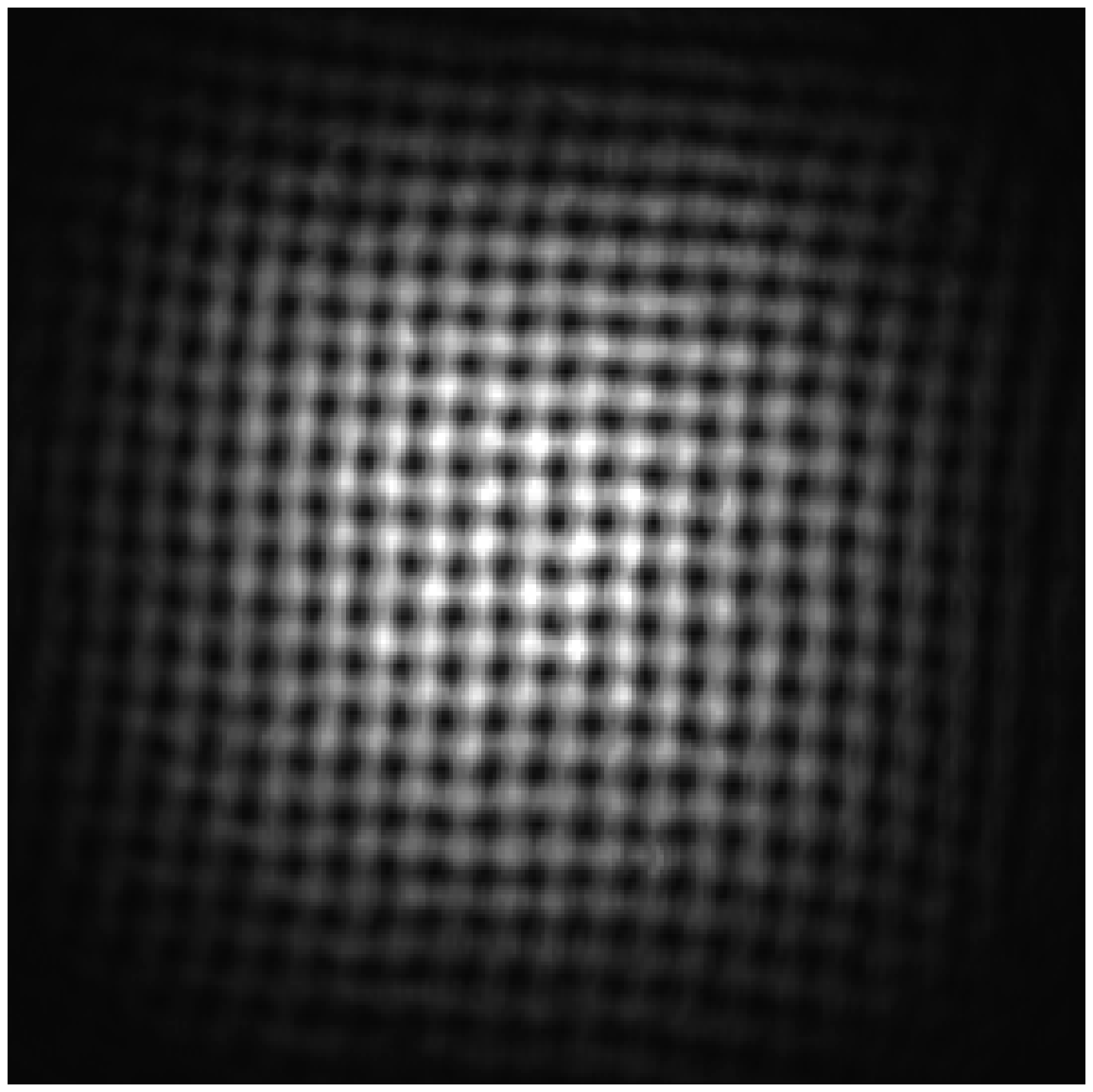}}
\subfigure[]{\label{fig:38}\includegraphics[scale=0.35]{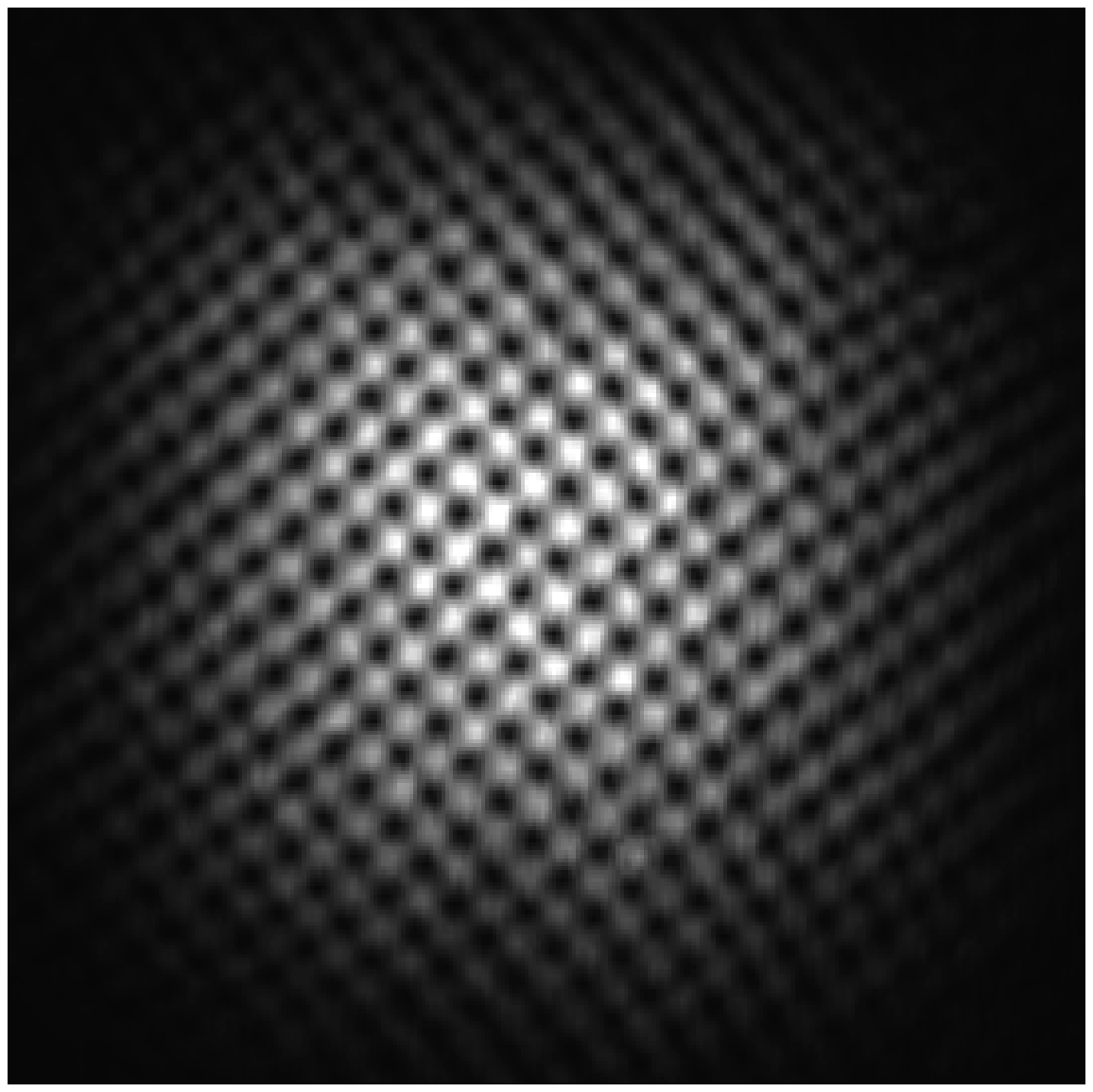}}
\subfigure[]{\label{fig:39}\includegraphics[scale=0.35]{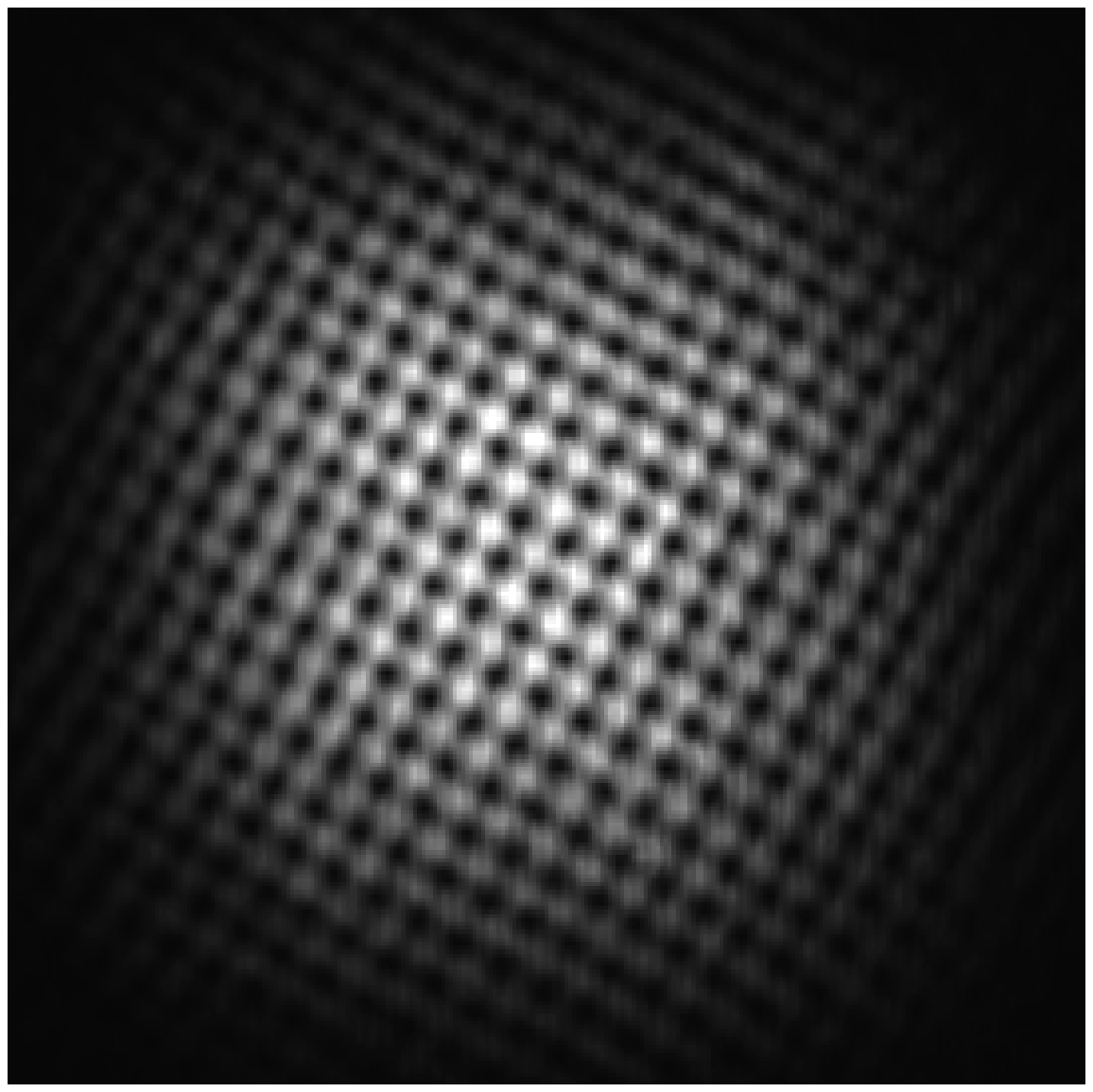}}
\subfigure[]{\label{fig:F_30_cir}\includegraphics[scale=0.38]{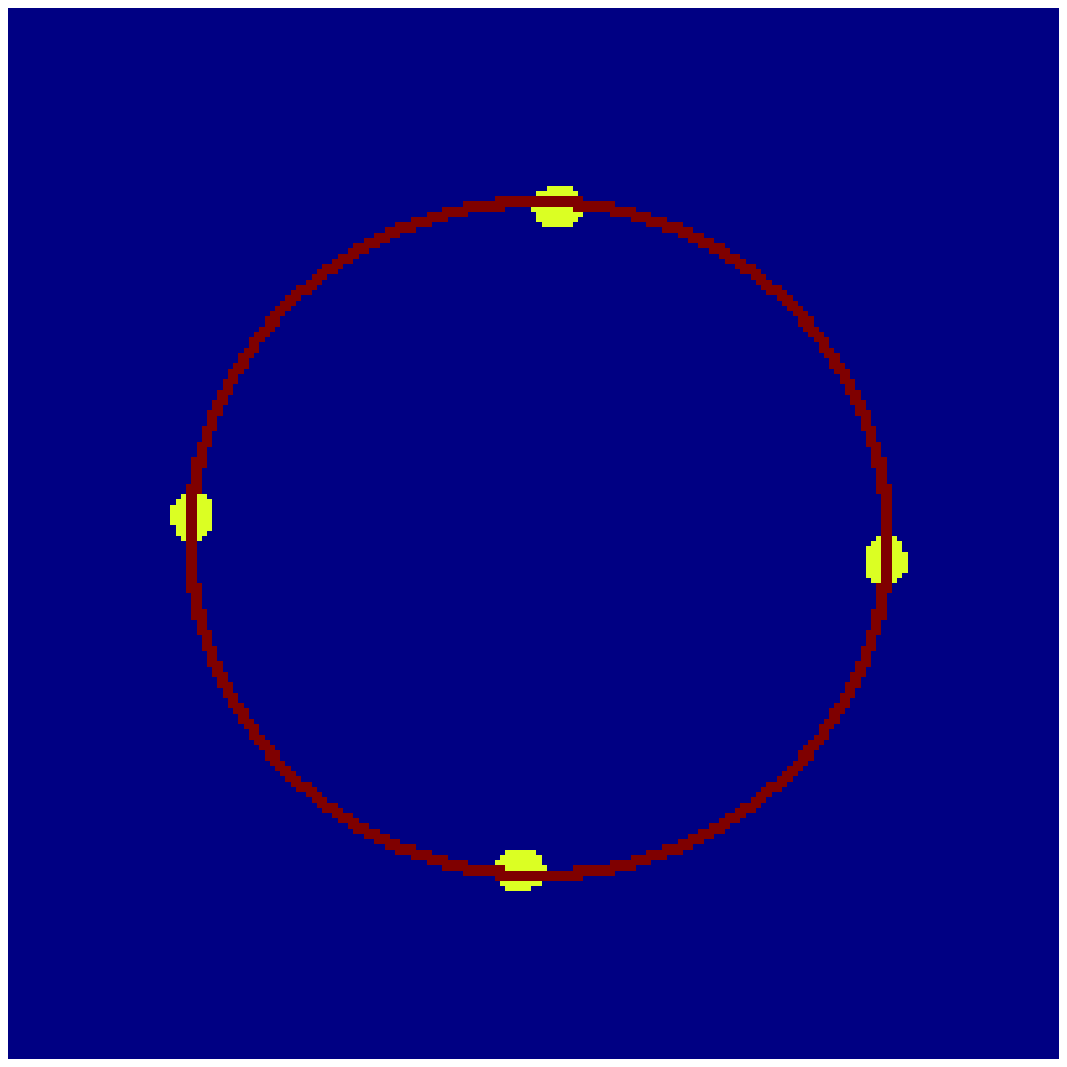}}
\subfigure[]{\label{fig:F_38_cir}\includegraphics[scale=0.38]{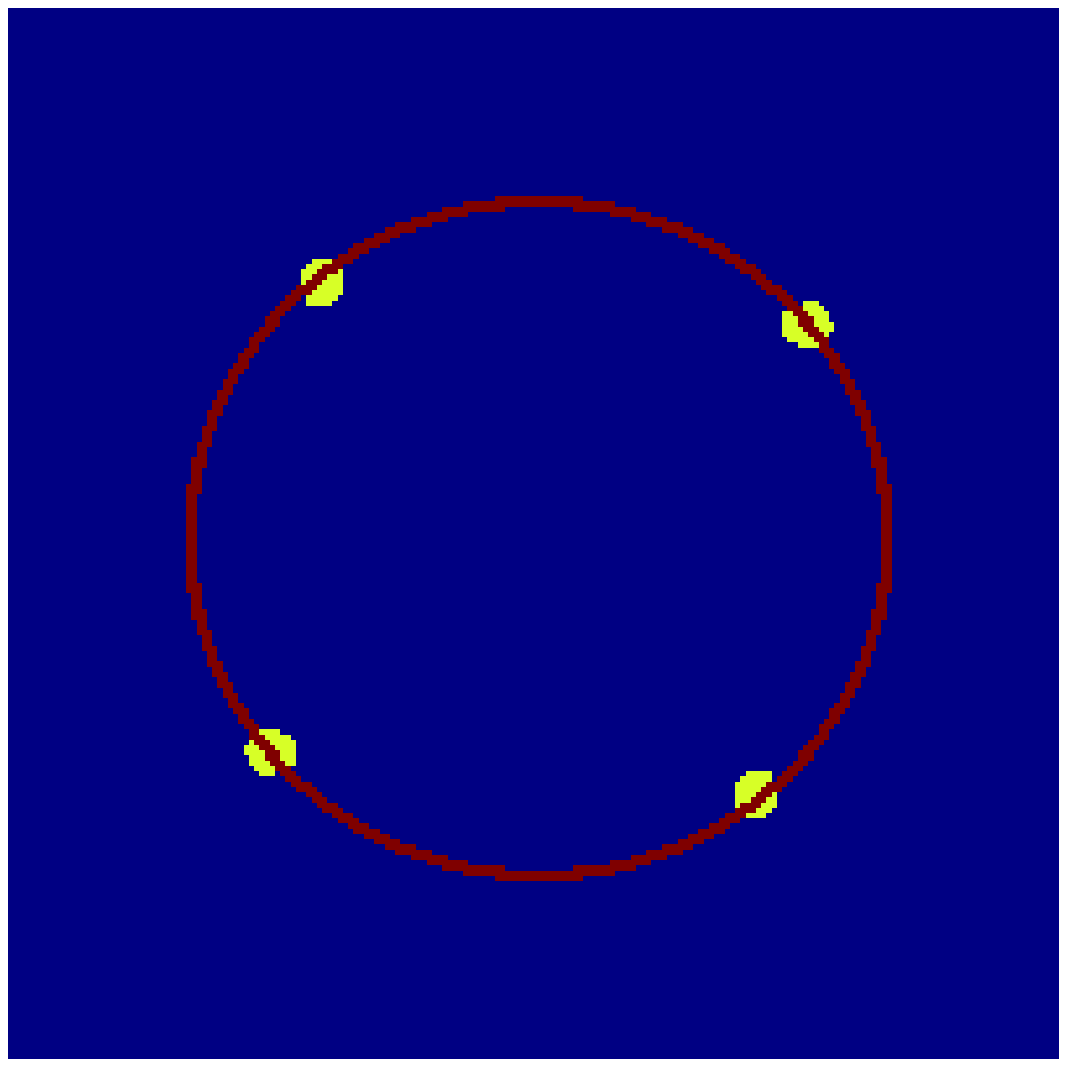}}
\subfigure[]{\label{fig:F_39_cir}\includegraphics[scale=0.38]{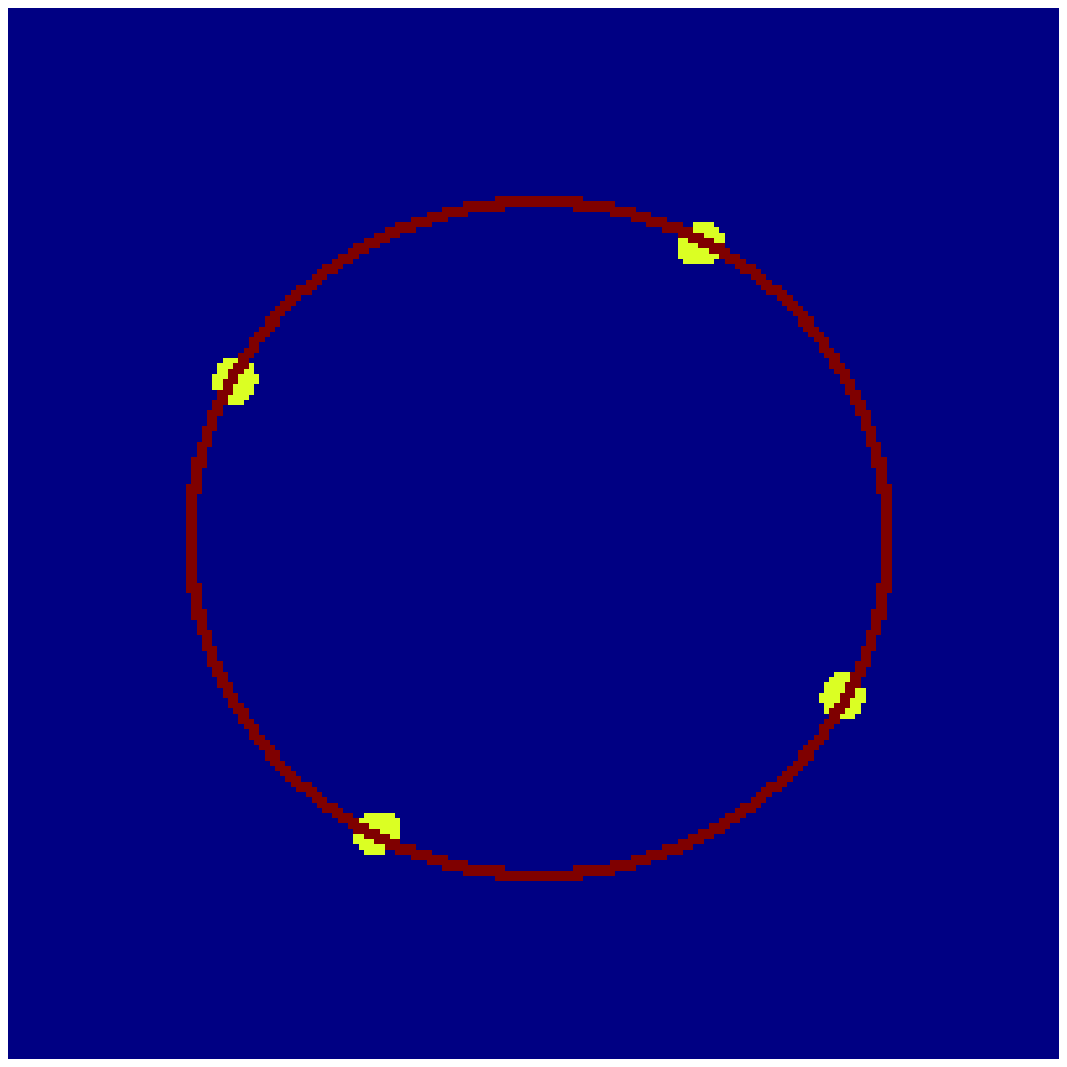}}
\caption{(a)-(c) (Multimedia online) Snapshots of the rotating lattice, taken while the lattice was rotating (Media 1). (d)-(f) Two dimensional 
Fourier transforms of Figs. (a)-(c) respectively.  The red circle has been added to the images to highlight that the periodicity of the lattice was 
constant as the lattice rotated.  The width of the peaks in the Fourier transforms represent the uncertainty in the lattice period, limited by the 
Fourier transform of the Gaussian envelope of the lattice.  The lattice spacing is 104(4)$\mu$m and the $1\slash e^2$ beam radius is 0.93mm.}
\label{fig:rotating snapshots}
\end{figure}

Figures \ref{fig:30}-\ref{fig:39} show snapshots of the optical lattice, taken while the optical lattice was rotating, indicating the lattice quality 
was not adversely affected by rotation (see also video online, showing the lattice rotating at 1Hz).  Figures \ref{fig:F_30_cir}-\ref{fig:F_39_cir} 
are the 2D Fourier transforms of \ref{fig:30}-\ref{fig:39} respectively.  The red circle passing through the peaks is the same in each image, and 
highlights that the periodicity of the lattice did not change as the lattice was rotated.

\begin{figure}[htbp]
\centering
\subfigure[]{\label{fig:0p8MHz}\includegraphics[scale=0.34]{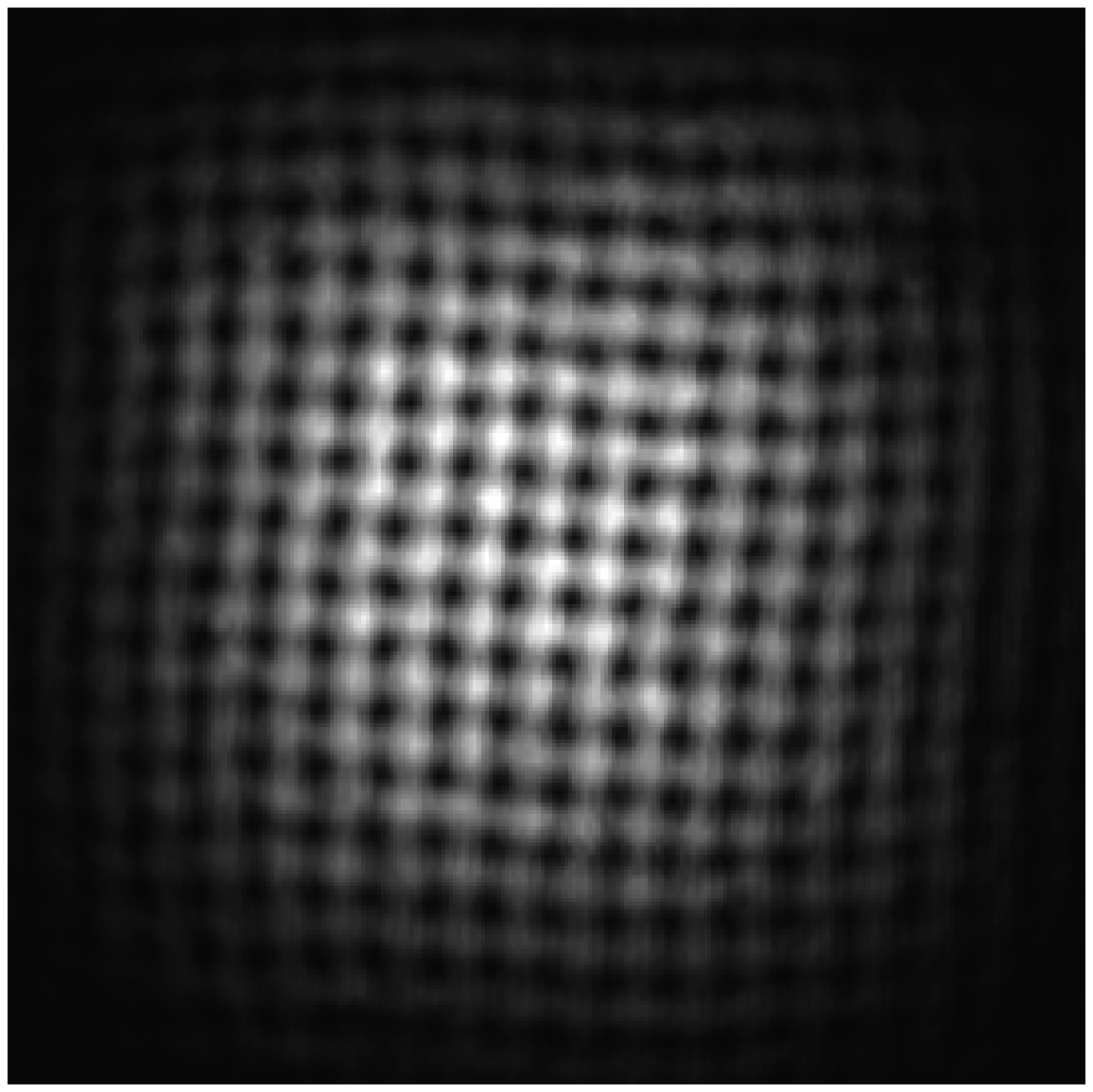}}
\subfigure[]{\label{fig:1p2MHz}\includegraphics[scale=0.34]{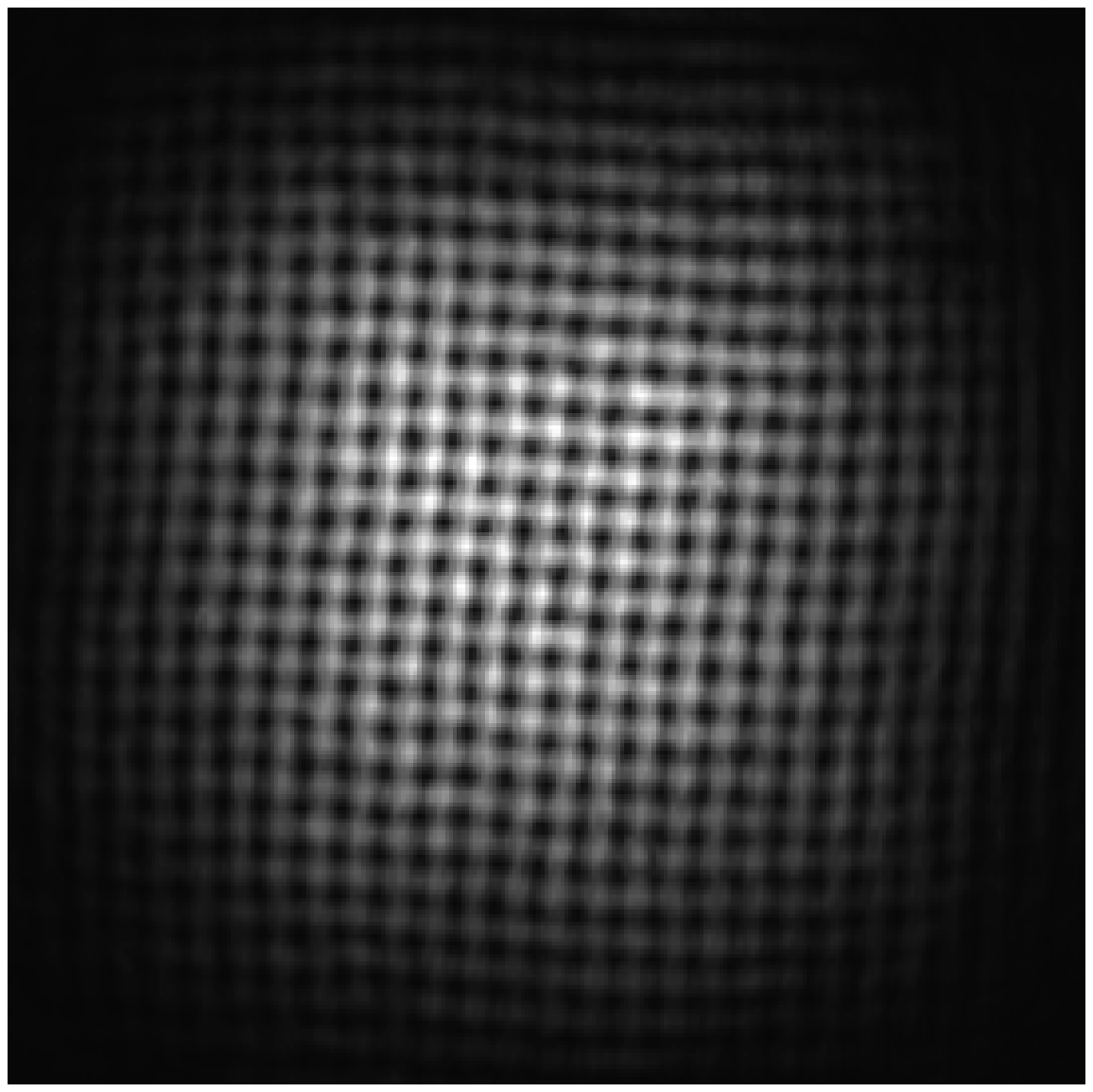}}
\subfigure[]{\label{fig:1p8MHz}\includegraphics[scale=0.34]{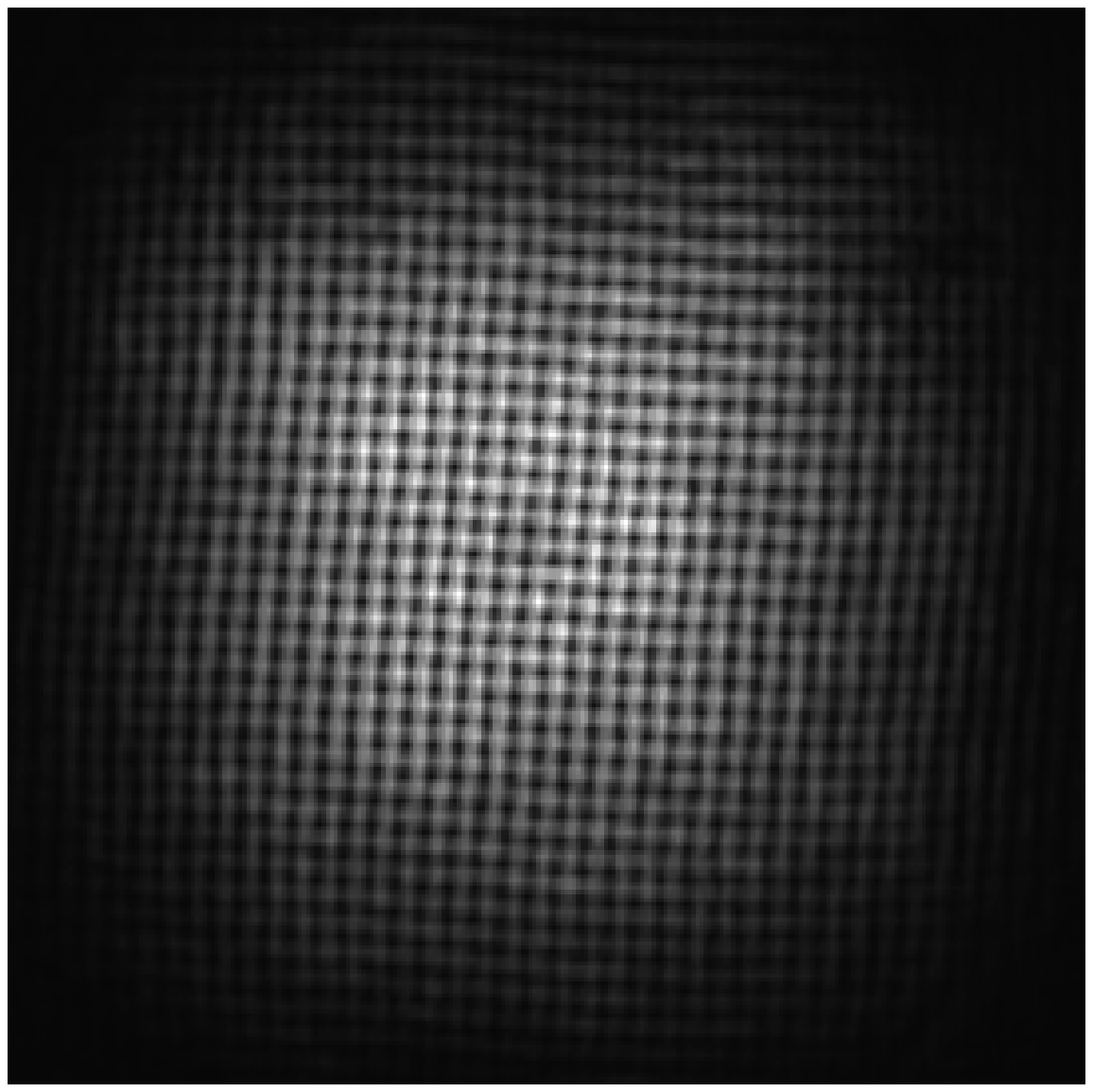}}
\caption{(a)-(c) (Multimedia online) Implementing a two-dimensional optical accordion lattice: pictures of the optical lattice at different 
periodicities (Media 2).  The lattice periodicities are: (a) 131(7)$\mu$m; (b) 87(3)$\mu$m; (c) 59(2)$\mu$m.  The $1\slash e^2$ beam radius is 0.93mm.  
The slight undulations that can be seen in the lattice are a result of wavefront aberration, and become less pronounced for smaller lattice spacings.}
\label{fig:accordion snapshots}
\end{figure}

Figure \ref{fig:accordion snapshots} (multimedia online) displays the two-dimensional accordion lattice.  As with rotation the lattice visibility can 
be seen to remain excellent as the periodicity of the lattice changes.

The images of Figs. \ref{fig:rotating snapshots} and \ref{fig:accordion snapshots} were taken by a CCD camera at the focus of objective lens L3.  The 
full scale of each image is $2.3\times 2.3$mm.  In order to achieve reasonable lattice periodicities that the camera could resolve for demonstrating 
the dynamic nature of the lattice we used a low N.A., 25cm focal length lens for L3.  For a cold atom experiment a high N.A. lens would be used, for 
example, we have since used an objective lens with an N.A. of 0.27 to load cold atoms into lattices with periodicities ranging from 1.8$\mu$m to 
18$\mu$m, details of which will be published in future work.

The advantages that the AOD-generated movement of the lattice beams brings to rotation, that is, fast and smooth rotation at a precise frequency, also 
apply to the two-dimensional accordion lattice.  The lattice spacing can be changed very quickly and precisely, with a maximum speed which is limited 
in our arrangement by the update rate of the digital synthesis system generating the rf signal for the AODs.  For example we could change the lattice 
spacing by a factor of 10 in under 100ms.

\section{Conclusion}
We have presented a versatile optical arrangement which can generate two-dimensional rotating and accordion lattices.  The scheme is based around an 
acousto-optic deflector, giving excellent, stable control over both rotation frequency and lattice periodicity as well as allowing a large range of 
rotation frequencies to be achieved.  Such a rotating lattice has important application in direct quantum simulation using ultracold atomic systems. 
The scheme presented here for dynamically controlling light intensity patterns could be applied to biophysics where intensity patterns generated by 
AODs have been used to optically sort particles flowing in water \cite{milne07}.

\section*{Acknowledgements}

This work is supported by EPSRC, QIP IRC and ESF.


\begin{thebibliography}{99}

\bibitem{lewenstein07} M. Lewenstein, A. Sanpera, V. Ahufinger, B. Damski, A. Sen(de), U. Sen, ``Ultracold atomic gases in optical lattices: Mimicking 
condensed matter physics and beyond,'' Adv. Phys. \textbf{56}, 243-379 (2007)

\bibitem{jain} J. K. Jain, {\it{Composite Fermions}} (OUP, 2007).

\bibitem{viefers08} S. Viefers, ``Quantum Hall physics in rotating Bose-Einstein condensates,'' J. Phys.:
Condens. Matter \textbf{20} 123202.1-123202.14 (2008)

\bibitem{sorenson05} A. S. Sorenson, E. Demler, M. D. Lukin, ``Fractional Quantum Hall States of Atoms in Optical Lattices,'' \prl \textbf{94}, 086803 
(2005)

\bibitem{palmer06} R.N. Palmer and D. Jaksch, ``High field fractional quantum Hall effect in optical lattices,'' \prl \textbf{96}, 180407 (2006)

\bibitem{bhat07} R. Bhat, M. Kr\"amer, J. Cooper, M. J. Holland, ``Hall effects in Bose-Einstein condensates in a rotating optical lattice,'' \pra 
\textbf{76}, 043601 (2007)

\bibitem{jaksch2003} D. Jaksch and P. Zoller, ``Creation of effective magnetic fields in optical lattices: the Hofstadter butterfly for cold neutral 
atoms,'' New J. Phys. \textbf{5}, 56.1-56.11 (2003).

\bibitem{polini05} M. Polini, R. Fazio, A. H. MacDonald and M. P. Tosi, ``Realization of Fully Frustrated Josephson Junction Arrays with Cold Atoms,'' 
\prl \textbf{95}, 010401 (2005)

\bibitem{kasamatsu08} K. Kasamatsu, ``Uniformly frustrated bosonic Josephson junction arrays,'' arXiv:0806.2012 

\bibitem{tung07} S. Tung, V. Schweikhard, and E. A. Cornell, ``Observation of Vortex Pinning in Bose-Einstein Condensates,'' \prl \textbf{97}, 240402 
(2006)

\bibitem{kasamatsu06} K. Kasamatsu, M. Tsubota, ``Dynamical vortex phases in a Bose-Einstein condensate driven by a rotating optical lattice,'' \prl 
\textbf{97}, 240404 (2006)

\bibitem{huckans06} ``Optical Lattices and Quantum degenerate $^{87}$Rb in reduced dimensions,'' John Howard Huckans, PhD thesis, University of 
Maryland (2006)

\bibitem{fallani05} L. Fallani, C. Fort, J. E. Lye, and M. Inguscio, ``Bose-Einstein condensate in an optical lattice with tunable spacing: transport 
and static properties,'' \opex \textbf{13}, 4303-4313 (2005)

\bibitem{li08} T. C. Li, H. Kelkar, D. Medellin, and M. G. Raizen, ``Real-time control of the periodicity of a standing wave: an optical accordion,'' 
\opex \textbf{16}, 5465-5470 (2008)

\bibitem{greiner02} M. Greiner, O. Mandel, T. Essingler, T. W. H\"ansch and I. Bloch, ``Quantum phase transition from a superfluid to a Mott insulator 
transition in a gas of ultracold atoms,'' Nature \textbf{415}, 39-44 (2002)

\bibitem{jaksch04} D. Jaksch, ``Optical Lattices, Ultracold Atoms and Quantum Information Processing,'' Contemp. Phys. \textbf{45}, 367-381 (2004). 

\bibitem{saleh} B.E.A Saleh and M.C. Teich, {\it{Fundamentals of Photonics}} (Wiley, 1991) pp. 136-139

\bibitem{born} M. Born and E. Wolf, {\it{Principles of Optics, 7th Edition}} (CUP, 1999) pp. 178-180

\bibitem{schlosser01} N. Schlosser, G. Reymond, I. Protsenko and P. Grangier, ``Sub-poissonian loading of single atoms in a microscopic dipole trap,'' 
Nature \textbf{411}, 1024-1027 (2001)

\bibitem{schnelle08} S. K. Schnelle, E. D. van Ooijen, M. J. Davis, N. R. Heckenberg and H. Rubinsztein-Dunlop, ``Versatile two-dimensional potentials 
for ultracold atoms,'' \opex \textbf{16}, 1405-1412 (2008)

\bibitem{milne07} G. Milne, D. Rhodes, M. MacDonald, and K. Dholakia, ``Fractionation of polydisperse colloid with acousto-optically generated 
potential energy landscapes,'' \ol \textbf{32}, 1144-1146 (2007)


\end{thebibliography}
\end{document}